\documentstyle[twoside,epsf]{article}

\newcommand{\ket}[1]{\left| {#1} \right>}
\newcommand{\C}{^C\!}

\catcode`\@=11
\long\def\@makefntext#1{
\protect\noindent \hbox to 3.2pt {\hskip-.9pt
$^{{\eightrm\@thefnmark}}$\hfil}#1\hfill}       

\def\thefootnote{\fnsymbol{footnote}}
\def\@makefnmark{\hbox to 0pt{$^{\@thefnmark}$\hss}}    

\def\ps@myheadings{\let\@mkboth\@gobbletwo
\def\@oddhead{\hbox{}
\rightmark\hfil\eightrm\thepage}
\def\@oddfoot{}\def\@evenhead{\eightrm\thepage\hfil
\leftmark\hbox{}}\def\@evenfoot{}
\def\sectionmark##1{}\def\subsectionmark##1{}}

\oddsidemargin=\evensidemargin
\renewcommand{\thefootnote}{\fnsymbol{footnote}}

\newcounter{sectionc}\newcounter{subsectionc}\newcounter{subsubsectionc}
\renewcommand{\section}[1] {\vspace{12pt}\addtocounter{sectionc}{1}
\setcounter{subsectionc}{0}\setcounter{subsubsectionc}{0}\noindent
    {\tenbf\thesectionc. #1}\par\vspace{5pt}}
\renewcommand{\subsection}[1] {\vspace{12pt}\addtocounter{subsectionc}{1}
\setcounter{subsubsectionc}{0}\noindent
{\bf\thesectionc.\thesubsectionc. {\kern1pt \bfit #1}}\par\vspace{5pt}}
\renewcommand{\subsubsection}[1] {\vspace{12pt}\addtocounter{subsubsectionc}{1}
    \noindent{\tenrm\thesectionc.\thesubsectionc.\thesubsubsectionc.
    {\kern1pt \tenit #1}}\par\vspace{5pt}}
\newcommand{\nonumsection}[1] {\vspace{12pt}\noindent{\tenbf #1}
    \par\vspace{5pt}}

\newcounter{appendixc}
\newcounter{subappendixc}[appendixc]
\newcounter{subsubappendixc}[subappendixc]
\renewcommand{\thesubappendixc}{\Alph{appendixc}.\arabic{subappendixc}}
\renewcommand{\thesubsubappendixc}
    {\Alph{appendixc}.\arabic{subappendixc}.\arabic{subsubappendixc}}

\renewcommand{\appendix}[1] {\vspace{12pt}
        \refstepcounter{appendixc}
        \setcounter{figure}{0}
        \setcounter{table}{0}
        \setcounter{lemma}{0}
        \setcounter{theorem}{0}
        \setcounter{corollary}{0}
        \setcounter{definition}{0}
        \setcounter{equation}{0}
        \renewcommand{\thefigure}{\Alph{appendixc}.\arabic{figure}}
        \renewcommand{\thetable}{\Alph{appendixc}.\arabic{table}}
        \renewcommand{\theappendixc}{\Alph{appendixc}}
        \renewcommand{\thelemma}{\Alph{appendixc}.\arabic{lemma}}
        \renewcommand{\thetheorem}{\Alph{appendixc}.\arabic{theorem}}
        \renewcommand{\thedefinition}{\Alph{appendixc}.\arabic{definition}}
        \renewcommand{\thecorollary}{\Alph{appendixc}.\arabic{corollary}}
        \renewcommand{\theequation}{\Alph{appendixc}.\arabic{equation}}
        \noindent{\tenbf Appendix \theappendixc #1}\par\vspace{5pt}}
\newcommand{\subappendix}[1] {\vspace{12pt}
        \refstepcounter{subappendixc}
        \noindent{\bf Appendix \thesubappendixc. {\kern1pt \bfit #1}}
    \par\vspace{5pt}}
\newcommand{\subsubappendix}[1] {\vspace{12pt}
        \refstepcounter{subsubappendixc}
        \noindent{\rm Appendix \thesubsubappendixc. {\kern1pt \tenit #1}}
    \par\vspace{5pt}}

\newcommand{\textlineskip}{\baselineskip=13pt}
\newcommand{\smalllineskip}{\baselineskip=10pt}

\newcommand{\copyrightheading}[1]
    {\vspace*{-2.5cm}\smalllineskip{\flushleft
    {\footnotesize Quantum Information and Computation, Vol.~1, No.~0 (2001) 000--000 #1}\\
    {\footnotesize \copyright\kern2pt Rinton Press}\\
     }}

\newcommand{\publisher}[2]{{\begin{center}\footnotesize\smalllineskip
    Received #1\\
    Revised #2
    \end{center}
    }}

\def\abstracts#1#2#3{{
    \centering{\begin{minipage}{4.5in}\footnotesize\baselineskip=10pt
    \parindent=0pt #1\par
    \parindent=15pt #2\par
    \parindent=15pt #3
    \end{minipage}}\par}}

\def\keywords#1{{
    \centering{\begin{minipage}{4.5in}\footnotesize\baselineskip=10pt
    {\footnotesize\it Keywords}\/: #1
     \end{minipage}}\par}}
\def\communicate#1{{
    \centering{\begin{minipage}{4.5in}\footnotesize\baselineskip=10pt
    {\footnotesize\it Communicated by}\/: #1
     \end{minipage}}\par}}

\renewenvironment{thebibliography}[1]
        {\frenchspacing
     \ninerm\baselineskip=11pt
         \begin{list}{\arabic{enumi}.}
        {\usecounter{enumi}\setlength{\parsep}{0pt}
     \setlength{\leftmargin 12.7pt}{\rightmargin 0pt}
         \setlength{\itemsep}{0pt} \settowidth
    {\labelwidth}{#1.}\sloppy}}{\end{list}}

\newcounter{itemlistc}
\newcounter{romanlistc}
\newcounter{alphlistc}
\newcounter{arabiclistc}

\newcommand{\fcaption}[1]{
        \refstepcounter{figure}
        \setbox\@tempboxa = \hbox{\footnotesize Fig.~\thefigure. #1}
        \ifdim \wd\@tempboxa > 5in
           {\begin{center}
        \parbox{5in}{\footnotesize\smalllineskip Fig.~\thefigure. #1}
            \end{center}}
        \else
             {\begin{center}
             {\footnotesize Fig.~\thefigure. #1}
              \end{center}}
        \fi}

\newcommand{\tcaption}[1]{
        \refstepcounter{table}
        \setbox\@tempboxa = \hbox{\footnotesize Table~\thetable. #1}
        \ifdim \wd\@tempboxa > 5in
           {\begin{center}
        \parbox{5in}{\footnotesize\smalllineskip Table~\thetable. #1}
            \end{center}}
        \else
             {\begin{center}
             {\footnotesize Table~\thetable. #1}
              \end{center}}
        \fi}

\def\pmb#1{\setbox0=\hbox{#1}
    \kern-.025em\copy0\kern-\wd0
    \kern.05em\copy0\kern-\wd0
    \kern-.025em\raise.0433em\box0}

\def\fnt#1#2{\footnotetext{\kern-.3em
    {$^{\mbox{\scriptsize #1}}$}{#2}}}

\def\fpage#1{\begingroup
\voffset=.3in
\thispagestyle{empty}\begin{table}[b]\centerline{\footnotesize #1}
    \end{table}\endgroup}

\def\runninghead#1#2{\pagestyle{myheadings}
\markboth{{\protect\footnotesize\it{\quad #1}}\hfill}
{\hfill{\protect\footnotesize\it{#2\quad}}}}
\font\tenrm=cmr10
\font\tenit=cmti10
\font\tenbf=cmbx10
\font\bfit=cmbxti10 at 10pt
\font\ninerm=cmr9

\font\eightrm=cmr8

\def\FigName{figure}%
\newbox\captionbox
\long\def\@makecaption#1#2{%
  \ifx\FigName\@captype
    \vskip\abovecaptionskip
    \setbox\tempbox\hbox{{\figurecaptionfont #1\hskip1em #2}}
    \ifdim\wd\tempbox< 28pc
    \centerline{\box\tempbox}
    \else
    {\figurecaptionfont #1\hskip1em #2\par}
\fi\else
    \setbox\tempbox\hbox{{\tablecaptionfont #1\hskip1em #2}}
    \ifdim\wd\tempbox< 28pc
    \centerline{\box\tempbox}
    \else
    {\tablecaptionfont #1\hskip1em #2\par}%
    \fi
 \vskip\belowcaptionskip
 \fi}
\InputIfFileExists{psfig.sty}
{\typeout{^^Jpsfig.sty inputed...ok}}{\typeout{^^JWarning: psfig.sty could be be found.^^J}}
\InputIfFileExists{epsfsafe.tex}
{\typeout{^^Jepsfsafe.tex inputed...ok}}
            {\typeout{^^JWarning: epsfsafe.tex could not be found.^^J}}
\InputIfFileExists{epsfig.sty}
{\typeout{^^Jepsfig.sty inputed...ok}}{\typeout{^^JWarning: epsfig.sty could not be found.^^J}}
\InputIfFileExists{epsf.sty}
{\typeout{^^Jepsf.sty inputed...ok}}{\typeout{^^JWarning: epsf.sty could not be found.^^J}}%
\def\fps@figure{tbp}
\def\ftype@figure{1}
\def\ext@figure{lof}
\def\fnum@figure{Fig.\ \thefigure}
\def\qed{\hbox{${\vcenter{\vbox{              
   \hrule height 0.4pt\hbox{\vrule width 0.4pt height 6pt
   \kern5pt\vrule width 0.4pt}\hrule height 0.4pt}}}$}}

\renewcommand{\thefootnote}{\fnsymbol{footnote}}  

\begin{document}
\setlength{\textheight}{8.0truein}    

\runninghead{Quantum Computer Architecture   $\ldots$}
            {Andrew M. Steane}

\normalsize\textlineskip
\thispagestyle{empty}
\setcounter{page}{1}

\copyrightheading{} 

\vspace*{0.88truein}

\fpage{1}
\centerline{\bf QUANTUM COMPUTER ARCHITECTURE}
\vspace*{0.035truein}
\centerline{\bf FOR FAST ENTROPY EXTRACTION}
\vspace*{0.37truein}
\centerline{\footnotesize
ANDREW M. STEANE}
\vspace*{0.015truein}
\centerline{\footnotesize\it Centre for Quantum Computation, Oxford University, Parks Road}
\baselineskip=10pt
\centerline{\footnotesize\it Oxford OX1 3PU, England}

\vspace*{0.225truein}
\publisher{(received date)}{(revised date)}

\vspace*{0.21truein} \abstracts{ If a quantum computer is
stabilized by fault-tolerant quantum error correction (QEC), then
most of its resources (qubits and operations) are dedicated to the
extraction of error information. Analysis of this process leads to
a set of central requirements for candidate computing devices, in
addition to the basic ones of stable qubits and controllable gates
and measurements. The logical structure of the extraction process
has a natural geometry and hierarchy of communication needs; a
computer whose physical architecture is designed to reflect this
will be able to tolerate the most noise. The relevant networks are
dominated by quantum information transport, therefore to assess a
computing device it is necessary to characterize its ability to
transport quantum information, in addition to assessing the
performance of conditional logic on nearest neighbours and the
passive stability of the memory. The transport distances involved
in QEC networks are estimated, and it is found that a device
relying on swap operations for information transport must have
those operations an order of magnitude more precise than the
controlled gates of a device which can transport information at
low cost.}{}{}

\vspace*{10pt}
\vspace*{3pt}
\communicate{to be filled by the Editorial}

\vspace*{1pt}\textlineskip  
\section{Introduction}           
\vspace*{-0.5pt}
\noindent

\setcounter{footnote}{0}
\renewcommand{\thefootnote}{\alph{footnote}}

Fault-tolerant quantum error correction (QEC) may be the best way
to stabilize the operation of a quantum computer. The physical
basis of QEC is an orchestrated flow of entropy from the
computer's qubits out to the environment. In thermodynamic terms,
heat is generated in the computer by the imperfection of its
operations and by noise sources; the computing qubits are then
coupled to ancilliary qubits which have been prepared in
low-entropy, i.e. `cold', states, where the coupling is arranged
to have the special property that most of the information content
of the heat, but none of the information content of the logical
computation, passes to the ancillas.

There are various ways to arrange the details of this process, the
main ones which have been considered are (1) extract parity check
information one bit at a time into groups of qubits prepared in (a
close approximation to) `cat' states \cite{96:Shor,96:DiVincenzo};
(2) extract several parity
checks simultaneously by using ancillas prepared in codeword
states of quantum codes \cite{97:SteaneA,99:SteaneB};
(3) use a quantum code of toric geometry
so that each parity check only involves local groups of bits of
finite size \cite{97:KitaevA}. It emerges that in all these methods the resources
required to achieve very stable operation are such that the great
majority of the qubits and operations of the computer have to be
dedicated to this extraction of error information; the forward
evolution of the logical computation is, as it were, a small
subsidiary process proceeding on the back of the main business of
stabilising the machinary. (For example, for each logic gate of
the logical algorithm, there may be $> 10^4$ in the QEC networks.)

The most basic physical requirements of a device which can perform
quantum computation are mostly self-evident (a set of stable
sytems to act as qubits, controllable coupling between these
systems, a means of measuring the final state); a useful more
detailed consideration of them has been provided by DiVincenzo
\cite{00:DiVincenzo}. However, in view of the fact that what a
quantum computer has to achieve is mostly QEC rather than a
general algorithm, there exists a further basic consideration for
the physical archicture: the physical device should be one which
is well-suited to the requirements of QEC.

This paper will consider what those requirements are, and hence
propose a set of desirable properties for quantum computing
devices. Apart from the obvious ones of high precision and speed,
the most important further properties which emerge are an emphasis
on the ease with which information can flow (controllably) in the
computer, a suitable geometrical structure of the quantum
communication pathways, and the differing requirements of qubits
which play differing roles in the computer.

\section{Logical structure of quantum error correction}
\noindent

The cooling rate offered by QEC is restricted by two main
limitations: the space-time size and structure of the network
required to extract error syndromes fault-tolerantly, and the
physical communication problem. The latter arises because qubits
are physical entities which have to occupy physical locations, so
that the reliability and speed with which one qubit can interact
with another must decrease as a function of the distance between
them. Also, measurements of qubits involve a quantum-to-classical
communication which can be slow compared to quantum-quantum
operations. The logical analysis of algorithms can completely
ignore such issues, but when considering the physical requirements
of the hardware, they are centrally important.

The communication problem been previously discussed for one type
of error-correcting code and physical device (concatenated code
and nearest-neighbour interactions
\cite{00:Gottesman,97:Aharonov}). Here I will make observations
which have a wider range of applicability. Some of these are
simple but not widely appreciated.

\begin{figure}[htbp]
\vspace*{13pt} \centerline{\psfig{file=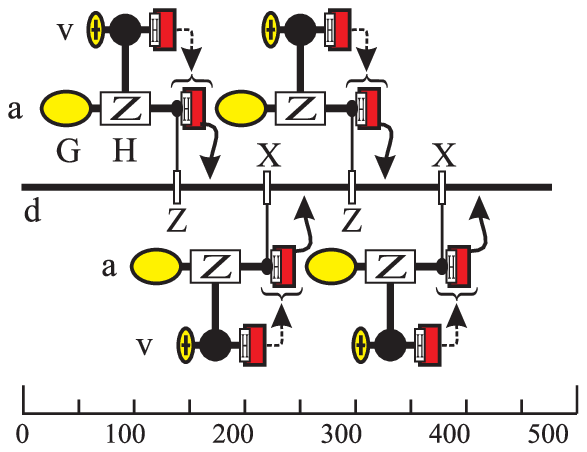,
width=4cm}} \vspace*{13pt} \fcaption{Network for two successive
recoveries, showing approximately the timing for the case of a
large efficient code of parameters $[[127,29,15]]$. d,a,v = 1
block of data, ancilla, verification bits respectively. $G,H$ =
networks implementing respectively the generator and check matrix
of the code. Ellipse=preparation network, red (grey)
rectangle=measurements, arrows=action conditional on measurement
results. The parallel preparation of many ancillas is not shown
here, but the need for repetition of syndrome extraction is taken
into account in the time allowed for the gates which couple to the
data. The recoveries can happen more frequently if further
ancillas are introduced. The time allowed for the $G$ and $H$
networks assumes the Latin rectangle method of
\protect\cite{02:SteaneA}.}
\end{figure}

The QEC networks extract error syndromes. The best way to do this
fault-tolerantly is either to use a code with a suitable topology
\cite{97:KitaevA} or to prepare encoded $\ket{0}_L$ states in
ancilla blocks, purify these by verification measurements, and
then couple each verified ancilla block to a data block to allow
error information to pass to the ancilla, before measuring all the
ancilla bits \cite{97:SteaneA,99:SteaneB,98:Preskill} (figure 1).
The latter method works for all CSS codes; I adopt it because it
allows measurements on encoded bits to be absorbed into the QEC
process, which is a useful tool in fault-tolerant processing
\cite{99:SteaneB}, because fault-tolerant universal sets of gates
can be constructed most simply, and therefore robustly, for CSS
codes \cite{96:Shor,98:GottesmanA}, and the class includes good
codes, i.e. ones with parameters $[[n,k,d]]$ such that the rate
$k/n$ and relative distance $d/n$ both remain finite as $n,k,d
\rightarrow \infty$ \cite{96:Calderbank,96:SteaneB}.

The syndrome extraction process under consideration has within
it a natural hierarchy of
communication needs. The qubits which must be coupled to one
another most frequently are those within each ancilla block, since
the network to prepare and verify an ancilla is large (thousands
of 2-bit gates); to extract a reliable syndrome several ancilla
blocks must be able to undergo controlled-not with a given data
block; finally the logical algorithm progresses when one data
block is coupled to another data block, but this happens
infrequently, approximately once per syndrome extraction. All
these statements concern the flow of {\em information} inside the
computer, they need bear no relation to the physical
structure of the computer, such as the physical locations of the
qubits. The information flow they describe is summarized by figure 2.

\begin{figure}[htbp]
\vspace*{13pt}
\centerline{\psfig{file=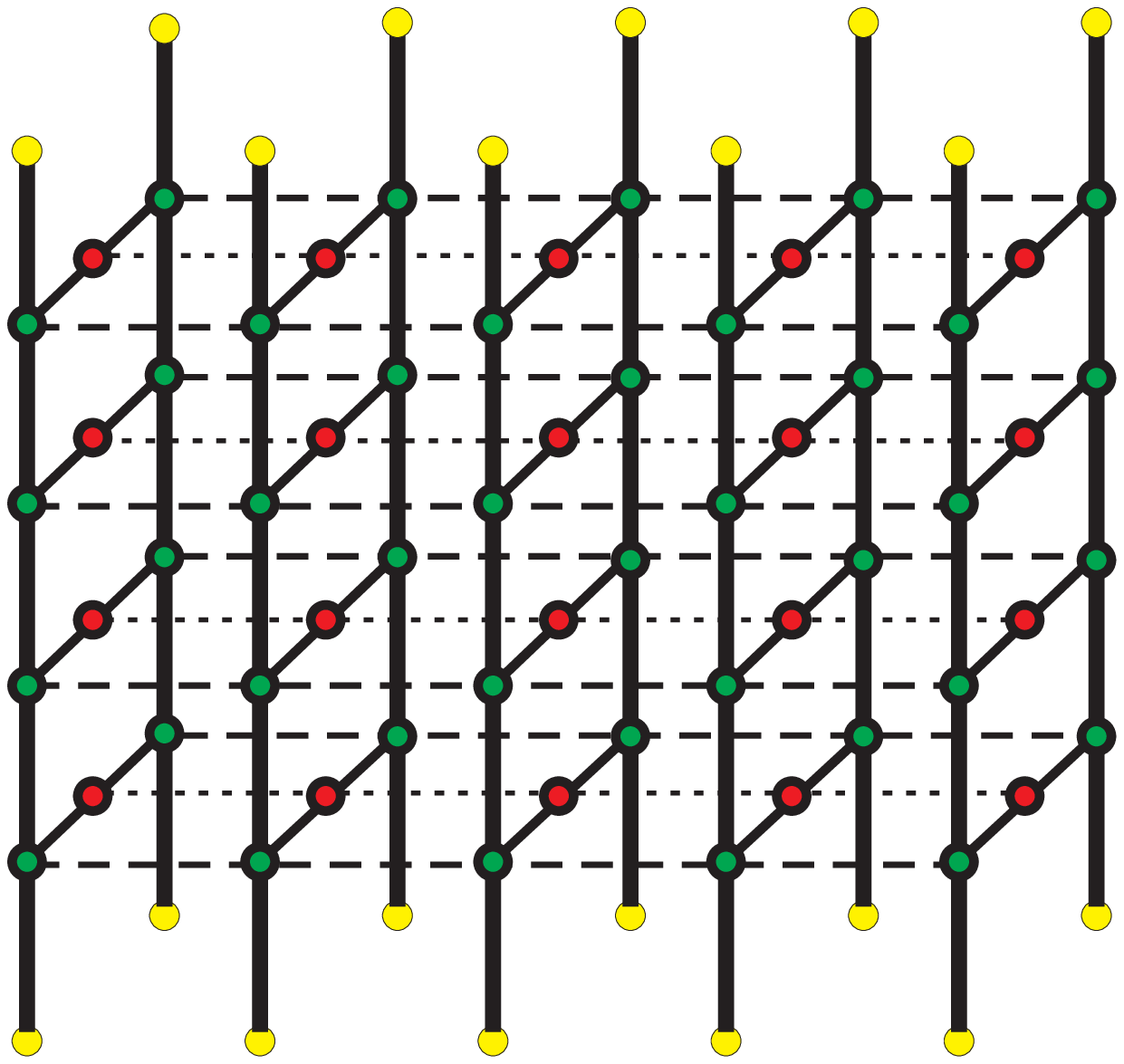, width=4cm}} 
\vspace*{13pt}
\fcaption{Information flow in a fault-tolerant quantum computer.
Each dot represents a physical qubit, and lines represent quantum
communication channels, with the information rate indicated by the
thickness of the lines. Red (dark gray) dots represent data bits,
green (gray) ancilla bits, and yellow (light) verification bits.
The ancilla and the
verification bits are measured each time a syndrome is extracted:
this further information flow to an external classical apparatus
takes place only at the external surfaces of the structure. Each
vertical column of data bits is one encoded block. The thick
vertical lines in ancillas represent the complex ancilla
preparation network, the horizontal lines between ancillas
represent transport of ancillas as needed to particular data
blocks, the dotted lines between data blocks represent
transversely applied logical gates in the algorithm. Different
data bits in the same block never communicate directly.}
\end{figure}

Using the diagram, a recovery of the computer is summarized as
follows. First a large number of gates operate on the front and
back surfaces of figure 2, in only the vertical direction.
Measurements are made of the verification bits at the top and
bottom, and then the good ancillas are transported horizontally so
that at least one good ancilla is adjacent to each data block. For
simplicity, let us assume the whole front surface is good, though
the argument will not depend on this. Then next the front surface
couples to the inner plane, i.e. the data blocks, by either
controlled-phase (for $X$ error correction) or controlled-not (for
$Z$ error correction). Then the front surface is measured. For the
blocks where a zero syndrome is deduced, nothing further happens,
while for those blocks where the syndrome is non-zero, the
pre-prepared (and as yet unused) good ancillas on the back surface
are coupled to the data to extract further syndromes so that a
majority vote can be taken. Finally, the result of the majority
vote is used to decide what corrective action to apply, if any,
and the relevant operation (one or more single-bit Pauli
rotations) is applied directly to the data qubits. The process is
repeated for the other type of error syndrome($X$ or $Z$), gates
are then applied between data bits of different blocks to evolve
the logical computation, and after this the recovery starts again.

\section{Physical implications}
\noindent

Moving information around is the main activity of the computer.
This process of moving information around is often summarized by
the shorthand notation of a vertical `gate' line extending over one or more
horizontal `qubit' lines in a quantum network diagram; figures
1 and 4 are examples. It is desirable to reduce the need for
information transport represented by these vertical lines, and
this can be done by careful network design (see below)
but the networks involved in verifying ancillas are not amenable to
being completely `untangled' in this way. That is, it is not
possible to arrange that the members of every pair of qubits involved in a
conditional gate are neighbours when their coupling is required, unless the
quantum information is shuffled around between the implementation of one controlled
gate and the next. This is especially true when the networks are made as
parallel as possible by using the Latin rectangle method described in
\cite{02:SteaneA}.

For each controlled-not ($\C X$) or controlled-phase ($\C Z$) gate, therefore,
qubits of information must be physically transported over relatively
large distances. The same is true when prepared ancillas are
transported horizontally in figure 2.

Recall that another large movement of information takes place
at the outer surfaces of figure 2, namely the measurement of the
ancillas and verification bits.
The timescale of this quantum-to-classical communication in the
measurement of the syndromes is important. Such measurement
is useful in order that the required substantial processing of the
syndromes can be done by reliable classical means, but the
time required for this measurement can be a significant fraction
of the total duration of the QEC process. Figure 1 gives an
example assuming that the time for a single-qubit
measurement is $t_m \simeq 25$ times that of a $\C X$ gate.This
is a reasonable model of ion trap processors but it is not yet
clear whether it is of other technologies.
If instead a quantum network is used to interpret the
syndrome in a unitary way, this further network will itself require
a time large compared to the time of a single elementary gate
and it must also allow qubits to be relaxed to $\ket{0}$.

The analysis above implies that the following are
primary considerations in designing a robust quantum computer
architecture: the physical structure of the computer should map
onto the information flow diagram, so that those physical bits
which have to communicate most frequently are able to do so
most reliably (and for preference at high rate); the physical
hardware should be well adapted to
information transport; qubits which never need to interact
directly, such as different physical bits in the same data block,
should be prevented from doing so, in order to minimize correlated
errors; fast measurements should be available for the verification
and ancilla qubits.

A physical hardware which realizes these properties in a natural
way is shown in figure 3. Although the information diagram of
figure 2 is 3-dimensional, the front-back dimension is much
smaller than the others (it could be 3 to 10 qubits, while the
others are hundreds or thousands of qubits for a large computer),
so in view of the difficulty of building controllable
3-dimensional structures, it makes sense to `squash' the diagram
into 2 dimensions while preserving the logical structure.

\begin{figure}[htbp]
\vspace*{13pt}
\centerline{\psfig{file=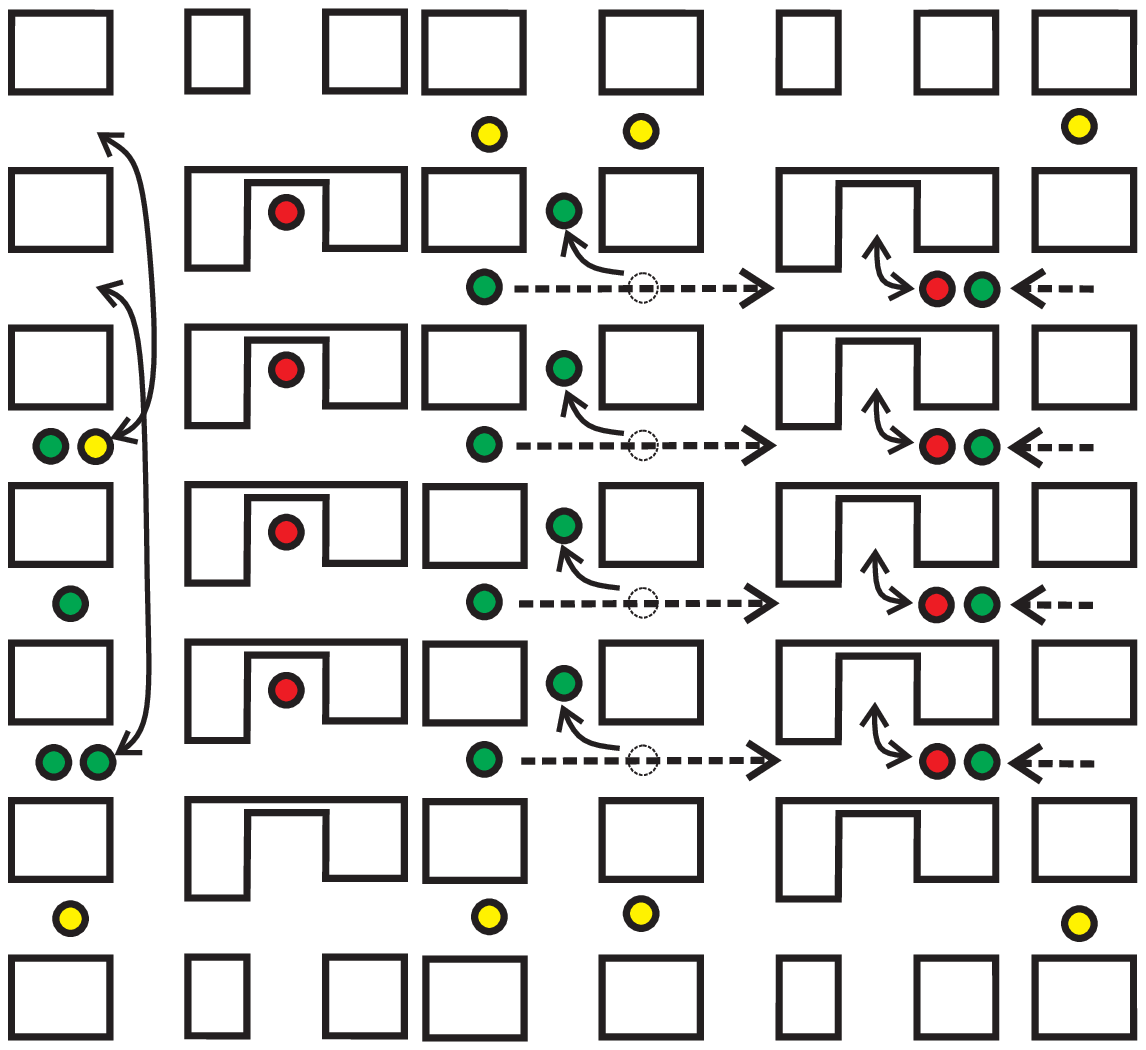, width=4cm}} 
\vspace*{13pt} \fcaption{Physical model of a quantum computer. The
dots represent qubits following the same colour code as figure 1.
The boxes represent segmented electrodes. The arrows represent
examples of information transport which are significant to
syndrome extraction and which can be performed simultaneously in
this model: controlled-not gates in the ancilla preparation,
horizontal transport of a prepared ancilla, and controlled-not
between an ancilla and a data block. In practice further ancillas
may be used (e.g. 4 rather than 2 per data block) to speed the
recovery rate.}
\end{figure}

Logic gates between non-neighbouring qubits are still necessary,
and there are many ways they can be accomplished, the main ones
being (1) multiple swap-operations between nearest neighbours, (2)
coupling to a physical entity such as a photon or phonon which
propagates between the bits, (3) transport of the physical entity
storing the qubit from one place to another. Teleportation may be
combined with any of these. Models (1) and (2) have been discussed
most often: (1) is a natural choice for several solid state
technologies, but requires many swaps for each desired logic gate,
therefore the swaps must be especially precise. (2) has advantages
associated with the speed of light and the insensitivity of
photons to electric field noise, but requires coupling between the
qubits and light which is hard to engineer in such a way as to
achieve all the communication represented in figure 2. Model (3)
is a natural choice for qubits stored in the fine or hyperfine
structure of movable ions or atoms, since the internal spin state
is very well preserved when atoms are transported in vacuum; it
may also be possible for electron spin qubits, with the electrons
transported inside a semiconductor. The transport model offers the
large degree of parallelism indicated in figure 3 in a simple way.

The thick vertical lines on figure 2 emphasize that most of the
processing is that required for ancilla preparation and
verification. One could envisage carrying out some of this using
qubits stored in degrees of freedom such as phonons or photons
which can interact simultaneously with several fixed qubits, in
order to prepare ancillas in fewer operations. However, the
following thermodynamic argument shows that not much can be gained
that way. The whole purpose of the ancilla preparation is to
prepare a physical entity in a state of low entropy, using
operations (quantum gates and communication) which are noisy. This
is possible because the {\em structure} of the network is itself a
highly ordered entity. This structure is guaranteed by the
classical device which controls the gates. The classical device is
assumed to be highly reliable at this level, i.e. it will never
generate gate operations in the wrong order, or on the wrong
qubits, etc. In order that the structure of the network can allow
sufficient negative entropy to flow into the preparation (along
with the undesirable but unavoidable entropy from the noise) the
network must be complicated and must involve a large number of
separate operations having uncorrelated noise.

\section{Typical gate distance}
\noindent

Suppose the speed and precision of a gate between qubits separated
by distance $s$ scales as $1 + s/D$; this is a reasonable estimate
for model (1) with $D=1$ and for model (3) with $D \gg 1$. If a
given computation requires a noise level below $\gamma$ per
controlled-not gate without taking communication costs into
account, and the gates are required between qubits separated on
average by $\bar{s}$, then the precision required per local gate
is of order $\gamma / \bar{s}$ and $\gamma$ in models (1) and (3)
respectively. If the precision of memory must be below $\epsilon$
per qubit-time-step without taking communication costs into
account, then it must be of order $\epsilon / \bar{s}$ and
$\epsilon$, in models (1) and (3) respectively.

In order to compare performances, and to interpret noise tolerance
calculations which do not take communication costs directly into
account, it is necessary to know the mean distance $\bar{s}$ for
the networks involved in error correction. I have calculated this
distance for two error correcting codes which have a special significance,
namely the $[[23,1,7]]$ Golay code and the $[[127,27,15]]$ BCH
code. The former has the best threshold behaviour when we take the
measurement time into account \cite{02:SteaneC}; the latter has an especially
good combination of space efficiency and noise
tolerance \cite{99:SteaneB, 02:SteaneC}; and the two
combined produce a powerful concatenated code.

The calculation assumed the qubits of any given ancilla and its
verification bits are layed out along a line, as in figs 2 and 3.
The calculation explicitly considered the verification network
(that labelled $H$ in fig. 1); the generation network $G$ is
similar but simpler so $H$ gives a better guide to the
requirements. The network was constructed by converting the check
matrix $H$ into the form $H = (I A)$ where $I$ is an identity
matrix and extracting $A$; a latin rectangle was constructed for
$A$; this rectangle produces the verification network as discussed
in \cite{02:SteaneA}. At this stage a network has been obtained
which gives the set of logical gates between logical qubits as a
function of time. The network was next converted into a
``shuttled" form, in which each two-bit gate is assumed to consist
of a transport of one of the two physical bits so that it becomes
adjacent to the other, with the intermediate bits shifted upwards
or downwards (as in a shift register), followed by the logical
gate between the now neighbouring bits. The rest of the network is
then adjusted to take account of the new bit locations. Thus the
physical locations of the logical bits continually change as each
successive gate is applied.

The details of the physical device will dictate what form these
bit displacements take. For example, the effect of multiple swap operations
is precisely the shift described, but if we transport bits as in
fig. 3 then it is not necessary to shift the intermediate bits. In
both cases the mean distance calculation works the same way, however, since it
computes the distances between bits measured in units of the number of intervening
bits.

The mean distance of the network is defined to be the separation of
the physical bits involved in a given two-bit gate in the ``shuttled"
network at the time gate is applied, (i.e. just before the transport operation),
averaged over all the gates of the network.

There is some flexibility in the construction of these networks.
For example, there are many latin rectangles of minimum alphabet
for the $A$ matrix, there is a choice of the ordering of the
logical bits when the ancilla state is first generated, and there
is a choice of which bit to transport in each gate. The network
was optimized by the following procedure. First a latin rectangle
of minimum alphabet was formed, and then adjusted so as to
minimize the maximum number of occurences of any given symbol in
the matrix without increasing the alphabet size. This minimizes
the number of gates which must operate in parallel given that a
network of fewest time steps is being used. Next the verification
network was constructed from the latin rectangle, and then
adjusted by use of a simulated annealing algorithm: the cost $c$
of a given change in the shuttled network was calculated and the
change introduced with probability $\exp(-c/T)$ where $T$ is a
temperature measure which was slowly reduced (changes with $c<0$
were always incorporated). The changes which were tried in this
way were random changes in the order of the bits of the ancilla at
the beginning of the network, and a change in the choice of which
bit to transport in each gate. The algorithm was designed to
reduce the maximum separation of any gate in the shuttled network,
and then among networks which agree on that to minimize the
$r.m.s.$ separation. To accomplish this, first $c = \Delta (j
\times {\rm max}(s))$ was used, where ${\rm max}(s)$ is the
maximum separation found in the shuttled network , and $j$ is the
number of gates having this separation; whenever this $c$ was zero
then $c = \Delta \left< s^2 \right>^{1/2}$ was used instead. The
simulated annealing algorithm was run through many cycles of
`heating' followed by slow `cooling' in order to search for a
global minimum.

\begin{figure}[htbp]
\vspace*{13pt}
\centerline{\psfig{file=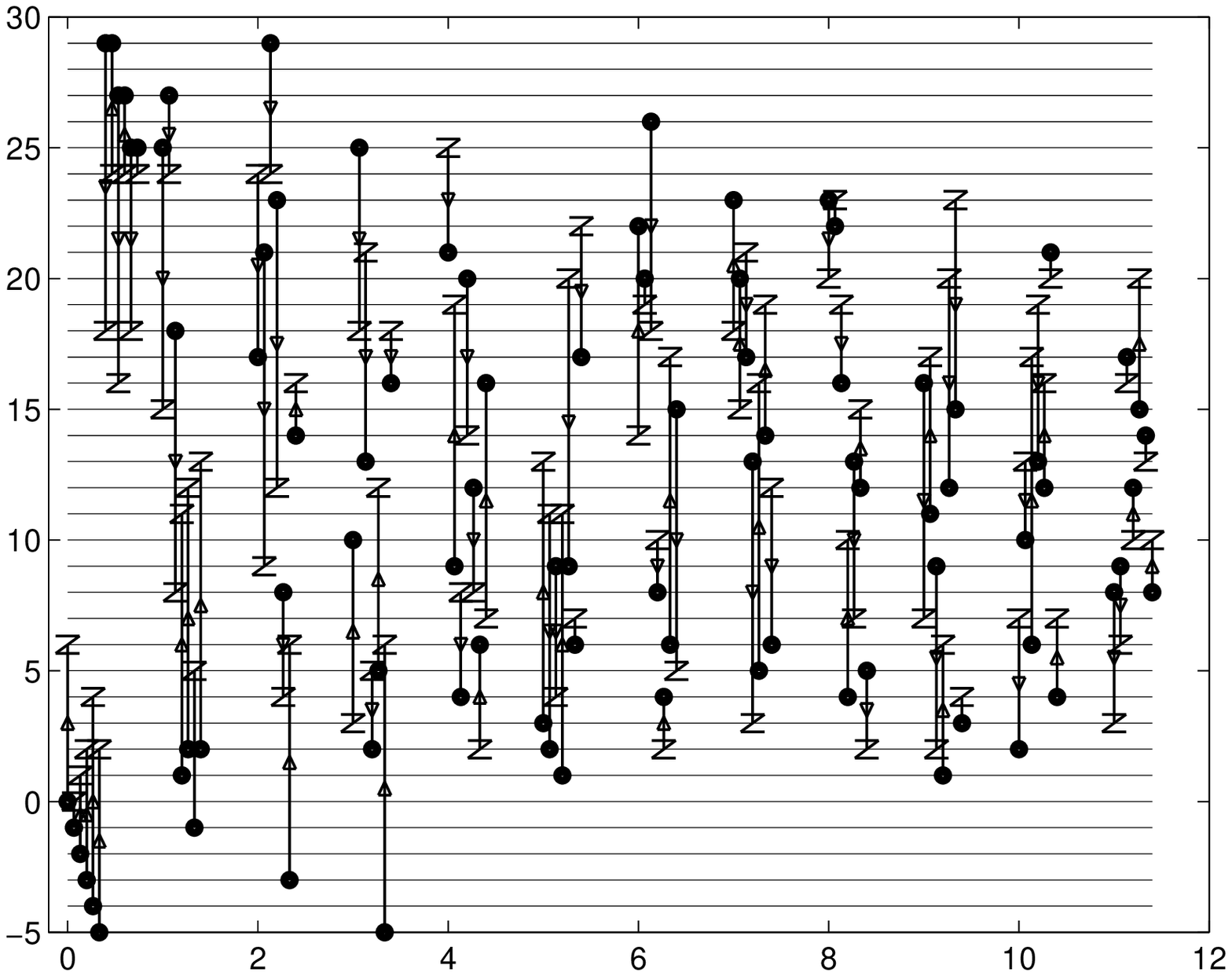, width=12cm}} 
\vspace*{13pt}
\fcaption{The network for the verification of ancillas in the Golay code.
The network shows physical qubits and is in ``shuttled" form. That is,
each gate symbol represents a shift operation followed by implementation of the
gate between nearest neighbours. The arrow in the gate line shows the direction
of movement of the gate bit which moved; the intermediate bits are displaced one position
in the opposite direction. The ancilla block is initially in the central 23 bits.
The horizontal axis shows time steps; the gates within each unit increment in time
can be simultaneous. The network is one of many logically equivalent ones, but has been
optimized for least time, then least parallelism, then least
maximum gate separation, then least r.m.s. gate separation.}
\end{figure}

The resulting network for the case of the Golay code is shown in
full in figure 4. This optimized network has a distribution of
gate distances with mean $\bar{s}_{\rm G} = 6$, median $5$ and
maximum $12$. 12 gates act in parallel at the start of the
network, and thereafter 7. The corresponding results for the
$[[127, 29 15]]$ BCH code were $\bar{s}_{\rm BCH} = 22$, median
$19$, maximum $72$. 78 gates act in parallel at the start of the
network, and thereafter 42. In both cases the mean distance can be
reduced slightly if more gates act in parallel at some steps, or
if a higher maximum distance is allowed. Also, further changes in
the structure of the network may allow slight reductions in
distance. A transport of the verification bits back out to the
side locations may be useful before they are measured. Such a
final transport would not influence the average distance of the
network significantly.

A fairly large computer could be stabilized by means of the BCH
code alone \cite{99:SteaneB,02:SteaneC}. In this case, and
assuming a physical arrangement similar to that of fig 3, the
results imply that a device where information transport is by
swaps between nearest neighbours (method (1)) will require those
swap operations to be approximately a factor 20 more precise than
the controlled-gate operations of a device where low-noise
transport (e.g. by method (3)) over a distance of $D > 20$ qubits
is available.

\begin{figure}[htbp]
\vspace*{13pt}
\centerline{\psfig{file=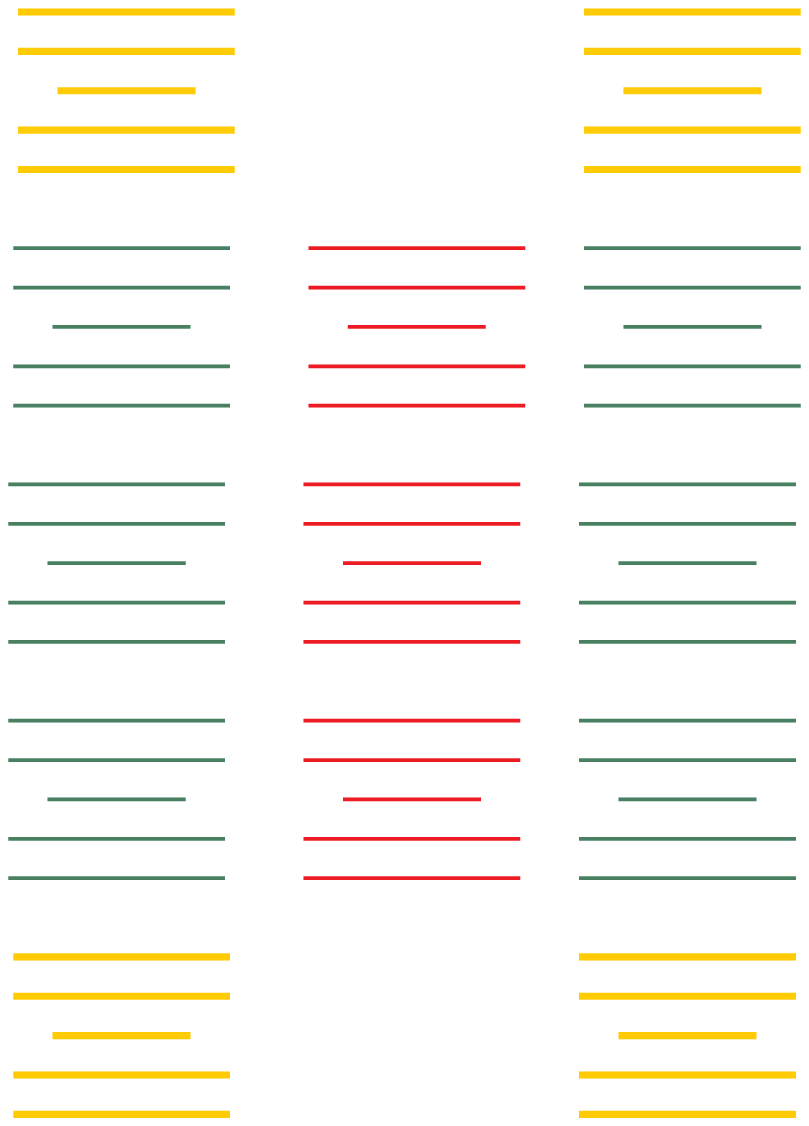, width=4cm}} 
\vspace*{13pt}
\fcaption{Physical layout for a computer based on a two-layer
concatenated code. Each horizontal line represents a block of 23 qubits encoded
in the inner code. A group of 5 such lines replaces each bit
in fig. 3. A single data block with a pair of ancillas is shown.}
\end{figure}

For very large quantum algorithms a concatenation of codes may be
necessary. In this case, each of the logical bits in figure 2 has
to be considered to be itself encoded and corrected; this adds a
further two dimensions to the information flow diagram for each
layer of concatenation. All these dimensions have to be compressed
into a maximum of 3 for the physical device; I will assume 2 here
for the layout of the qubits, the 3rd can be used for implementing
gates and readout. The resulting structure could be for example as
in figure 5. The inner, Golay, code works on the 23-bit inner
blocks and its operation is dominated by the ancilla preparation
and the gates between ancilla and data, therefore its noise
threshold will be a factor $f$ lower for the swap operations in
method (1) than for the controlled-gate operations in method (3),
with $f$ in the range approximately 3 to $\bar{s}_{\rm G} = 6$.
For the outer, BCH, code, it is seen from fig. 5 that vertical
distances (preparation and verification of ancillas) are
multiplied by a factor 5, compared to the bare BCH code, if 4
ancillas per block are adopted for the inner code (as in fig. 5,
this speeds the inner recovery time) or by a factor 3 if 2
ancillas per block are adopted. This gives a mean separation
between $3 \bar{s}_{\rm BCH} = 66$ and $5 \bar{s}_{\rm BCH} = 110$
for the vertical network. Horizontal distances for the transport
of ancillas and coupling to data are multiplied by 23, making a
mean horizontal distance of order 50 if we assume the required
ancilla transport distance at the outer level to be 2 on average.

Therefore if transport is by swap operations, the noise of these
must be small enough to reduce the swap gate noise of the outer
code by a factor in the range approximately 60 to 110, compared to
the case of transport at no cost. Since the inner code is 3-error
correcting, a reduction in physical gate noise by a factor
$110^{1/4} \simeq 3.2$ suffices. This is in addition to the factor
3 to 6 already mentioned, making 10 to 20 overall.

In model (3) the transport cost in the concatenated code is small
if $D > 40$ since then it is negligible for the inner recoveries,
and the noise associated with the vertical transport distances up
to approximately $3 D$ only causes a small fractional change in
the noise accumulating in the recovery networks for the Golay
code. This is because most of the noise to be corrected by each
inner recovery network then remains that associated with the inner
network itself.

I conclude that for this concatenated code, the swap operations in
model (1) must be an order of magnitude more precise than the
controlled-gates in model (3), assuming the transport
distance-scale $D$ in model (3) satisfies $D > 40$.

To conclude overall, transportation of information is the main
activity of a quantum computer stabilized by fault-tolerant QEC.
This flow has a natural geometry, and the physical computer should
be designed to reflect this geometry. The information flow needs
to be fast within the ancillas, and moderately fast between
ancillas and data, and between ancillas and a classical measuring
device, but can be relatively slow elsewhere. The size and
complexity of the set of operations to prepare ancillas cannot be
further reduced unless the individual operations are made more
precise, since otherwise insufficient entropy will be extracted.
Mean transport distances for optimized networks to implement two
important error correcting codes have been calculated. A computer
based on physical qubits which can themselves be easily
transported, as well as interact for gate operations, is
attractive for achieving the right kind of information flow.

\nonumsection{Acknowledgements}
\noindent

This work was supported by the EPSRC and the Research Training and
Development and Human Potential Programs of the European Union.

\nonumsection{References}
\noindent


\begin{thebibliography}{10}

\bibitem{96:Shor}
P.~W. Shor~(1996), {\em Fault-tolerant quantum computation}, in
{\em Proc.
  35th Annual Symposium on Fundamentals of Computer Science}, (Los Alamitos),
  pp.~56--65, IEEE Press.
\newblock quant-ph/9605011.

\bibitem{96:DiVincenzo}
D.~P. DiVincenzo and P.~W. Shor~(1996), {\em Fault-tolerant error
correction
  with efficient quantum codes}, Phys. Rev. Lett., 77, pp.~3260--3263.

\bibitem{97:SteaneA}
A.~M. Steane~(1997), {\em Active stabilisation, quantum
computation, and
  quantum state sythesis}, Phys. Rev. Lett., 78, pp.~2252--2255.
\newblock quant-ph/9608026.

\bibitem{99:SteaneB}
A.~M. Steane~(1999), {\em Efficient fault-tolerant quantum
computing},
  Nature, 399, pp.~124--126.
\newblock quant-ph/9809054.

\bibitem{97:KitaevA}
A.~Y. Kitaev~(1997), {\em Quantum error correction with imperfect
gates}, in
  {\em Quantum Communication, Computing and Measurement (Proc. 3rd Int. Conf.
  of Quantum Communication and Measurement)}, (New York), pp.~181--188, Plenum
  Press.

\bibitem{00:DiVincenzo}
D.~P. DiVincenzo~(2000), {\em The physical implementation of
quantum
  computation}, Fortschritte der Physik, 48, pp.~771--783.

\bibitem{00:Gottesman}
D.~Gottesman~(2000), {\em Fault-tolerant quantum computation with
local
  gates}, Journal of Modern Optics, 47, pp.~333--45.
\newblock quant-ph/9903099.

\bibitem{97:Aharonov}
D.~Aharonov and M.~Ben-Or~(1997), {\em Fault-tolerant quantum
computation
  with constant error rate}, in {\em Proc. 29th Annual ACM Symposium on Theory
  of Computing (STOC)}, p.~176, ACM Press.
\newblock quant-ph/9611025 and 9906129.

\bibitem{02:SteaneA}
A.~M. Steane~(2002), {\em A fast fault-tolerant filter for quantum
  codewords}, submitted for publication.
\newblock quant-ph/0202036.

\bibitem{98:Preskill}
J.~Preskill~(1998), {\em Reliable quantum computers}, Proc. R.
Soc. Lond. A,
  454, pp.~385--410.

\bibitem{98:GottesmanA}
D.~Gottesman~(1998), {\em A theory of fault-tolerant quantum
computation},
  Physical Review A, 57, pp.~127--137.
\newblock quant-ph/9807006.

\bibitem{96:Calderbank}
A.~R. Calderbank and P.~W. Shor~(1996), {\em Good quantum
error--correcting
  codes exist}, Phys. Rev. A, 54, pp.~1098--1105.

\bibitem{96:SteaneB}
A.~M. Steane~(1996), {\em Multiple particle interference and
quantum error
  correction}, Proc. Roy. Soc. Lond. A, 452, pp.~2551--2577.

\bibitem{02:SteaneC}
A.~M. Steane~(2002), {\em Overhead and noise threshold of
fault-tolerant
  quantum error correction}, in preparation.

\end{thebibliography}

\end{document}